\documentclass[%
 twocolumn,
 amsmath,amssymb,
 aps,
pra,
]{revtex4}

\usepackage{amsmath,braket,bm}
\usepackage{graphicx}
\usepackage{dcolumn}
\usepackage{bm}

\usepackage{color}

\usepackage{hyperref}
\hypersetup{colorlinks=true, linkcolor=blue, citecolor=red, urlcolor=blue}

\usepackage{physics}

\newcommand*{\Bellop}{\Pi_{\text{CHSH}}} 
\newcommand*{\DoubleBlochState}{\psi_{\theta,\phi}}
\newcommand*{\EnsembleForFirst}{\rho_{\theta,\phi}}
\newcommand*{\relu}{\sigma_{RL}}
\newcommand*{\identity}{I}
\newcommand*{\CHSHML}{CHSH$_{\text{ml}}$}

\begin{document}


\title{Transforming Bell's Inequalities into State Classifiers with Machine Learning}%

\author{Yue-Chi Ma}
 \affiliation{Center for Quantum Information, Institute for Interdisciplinary Information Sciences, Tsinghua University, Beijing,100084, China}
\author{Man-Hong Yung}%
\email{yung@sustc.edu.cn}
\affiliation{Institute for Quantum Science and Engineering and Department of Physics, Southern University of Science and Technology, Shenzhen 518055, China}
\affiliation{Center for Quantum Information, Institute for Interdisciplinary Information Sciences, Tsinghua University, Beijing,100084, China}


\begin{abstract}
Quantum information science has profoundly changed the ways we understand, store, and process information. A major challenge in this field is to look for an efficient means for classifying quantum state. 
For instance, one may want to determine if a given quantum state is entangled or not. However, the process of a complete characterization of quantum states, known as quantum state tomography, is a resource-consuming operation in general. An attractive proposal would be the use of Bell's inequalities as an entanglement witness, where only partial information of the quantum state is needed. 
The problem is that entanglement is necessary but not sufficient for violating Bell's inequalities, making it an unreliable state classifier.
Here we aim at solving this problem by the methods of machine learning.
More precisely, given a family of quantum states, we randomly picked a subset of it to construct a quantum-state classifier, accepting only partial information of each quantum state.
Our results indicated that these transformed Bell-type inequalities can perform significantly better than the original Bell's inequalities in classifying entangled states.
We further extended our analysis to three-qubit and four-qubit systems, performing classification of quantum states into multiple species. These results demonstrate how the tools in machine learning can be applied to solving problems in quantum information science.


\end{abstract}

\maketitle


\section{\label{sec:introduction}Introduction}
Quantum machine learning is an emerging field of research in the intersection between quantum physics and machine learning, which has profoundly changed the way we interact with data. It represents a new paradigm of processing information, which, at the fundamental level, is still governed by the laws of quantum mechanics. In addition, there is also a real ``demand" of using advanced data-processing techniques for gate-fidelity benchmarking and data analysis  for the state-of-the art quantum experiments. Therefore, understanding the connection between quantum information science and machine learning is a matter of great fundamental and practical interest.


In general, there are many ways where research in quantum machine learning has become fruitful. One way is to design quantum algorithms to speed up classical machine learning~\cite{lloyd2014quantum,rebentrost2014quantum,lloyd2013quantum,dunjko2016quantum,Schuld2017}. For example, quantum extensions of the principal component analysis (PCA)~\cite{lloyd2014quantum} and support vector machines (SVM)~\cite{rebentrost2014quantum,li2015experimental} have been invented. Furthermore, {quantum algorithms~\cite{lloyd2013quantum,cai2015entanglement,Schuld2017} are capable to accelerate some distance-based problems exponentially}. 

On the other hand, the other approach in quantum machine learning is to apply machine learning methods to study problems in quantum physics and quantum information science. In particular, classical machine-learning methods~\cite{Chapman2016,Hentschel2010} have been applied to many-body~\cite{carleo2016solving,carrasquilla2016machine,Levine2017}, superconducting~\cite{Magesan2015}, bosonic~\cite{Wang2016} and electronic~\cite{mills2017deep} systems. Furthermore, machine-learning can also be applied to the problem of quantum-Hamiltonian learning~\cite{wiebe2014hamiltonian,wang2017experimental}. Beyond quantum information science, machine learning also finds applications in particle physics~\cite{Chiappetta1994}, electronic structure of molecules~\cite{Rupp2015}, and gravitational physics~\cite{Biswas2013}. 

In this work, we are interested in applications of machine learning to the problem of quantum-state classification~\cite{Guta2010}, which is a generalization of the pattern recognition in learning theory. In the classical setting of pattern recognition, we are given a training set $\mathcal S$ containing paired values, ${\mathcal S}= \left\{ {({x_1},{y_1}),({x_2},{y_2}),({x_3},{y_3}),...} \right\}$, where $x_i$ is a data point and $y_i \in \{0,1\}$ is a pre-determined label for $x_i$. Furthermore, there are many classical methods in machine learning inspired by ideas in physics~\cite{ackley1985learning} and quantum information~\cite{wiebe2015quantum,sergioliquantum}. 

Based on the training set, the problem of pattern recognition is to construct a low-error classifier (or predictor), in the form of a function, $f:x \to y$, for predicting the labels of new data. The quantum extension of this problem is to replace the data points $x_i$ with density matrices of quantum states $\rho_i$, i.e., ${x_i} \to {\rho _i}$. The challenge is that obtaining full information for a given quantum states becomes resource consuming as the number of qubits increases.

Instead of full information (e.g. from quantum tomography), we aim at constructing a set of quantum-state classifiers which can reliably output a correct label of a given quantum state in an ensemble, using only partial information (i.e., a few observables) about the state. Our strategy is motivated by the development of Bell's inequalities, which was originally used to exclude incompatible classical theories from a few measurement results performed non-locally. 

In fact, there are challenges in using of Bell's inequalities for the purpose of state classification. Quantum mechanically, it is well-known that entanglement is necessary for violating, e.g., the CHSH (Clauser-Horne-Shimony-Holt) inequality (see also Eq.~(\ref{CHSH_original})). However, entanglement is not sufficient, meaning that there are many entangled states not violating the CHSH inequality, which makes it an unreliable state classifier for detecting quantum entanglement. 

Our strategy is to ``transform" Bell's inequalities into a reliable state classifier. However, the non-locality aspect of Bell's inequalities is not relevant to the construction of our quantum state classifiers, although we can follow the same experimental setting for an implementation of our proposal.

Here the transformation involves two levels. First, we ask the following question: ``{\it given the same measurement setting, is it possible to optimize the coefficients of the CHSH inequality for a better performance, compared with the values $(1,-1,1,1,2)$ employed in the standard CHSH inequality (see Eq.~(\ref{CHSH_original}))?}" We shall see that that the answer to this question is positive. This optimization is linear, in the sense that it gives an optimization function containing a linear combination of the observables as input. 

In the second level, instead of linear optimization, we include hidden layers in a non-linear optimization process and at the same time allow the measurement angles to be varied randomly. We found that in this way, the performance of the classifier can be enhanced significantly, relative to the first level. This method is then applied to several different scenarios of quantum state classification. Before we go into the details, we provide an overview and summary of the main results below.

\subsection{Overview and main results}

In this work, we present an application of supervised machine learning to the problem of classifying quantum states, where we construct quantum-state classifiers form some training set of quantum states. Here a classifier for quantum states output a ``label", for example, {\sf entangled} or {\sf unentangled}, for any quantum state, by sending partial (or fully tomographic in some cases) measurement data of the quantum state as input. 

More specifically, we consider the scenarios where quantum states are distributed to different parties through an noise channel characterized by some unknown parameter. The parties are given the opportunity to test the channel through a set of testing states, which corresponds to the training phase of machine learning. At the end, the parties are given an non-linear function optimized for the purpose of state classification, where only partial information is required for testing new quantum states beyond the training set.

Our non-linear quantum-state classifier is constructed by a technique in machine learning known as ``artificial multilayer perceptron"~\cite{MacKay2005}, which {is a network composed of several layers, where information flows from input layer, through hidden layer, and finally to the output layer.} 

The input layer contains the information about the quantum state, where the expectation value of certain observables are taken as the elements of a vector $\vec{x}$. The hidden layer contains another vector  $\vec{x_1}$, which is constructed through the relation, 
\begin{equation}
\vec{x}_1 = \relu(W_1\vec{x}+\vec{w}_{01}) \ .
\end{equation}
Here $W_1$, $\vec{w}_{01}$ are initialized uniformly and optimized through the learning process. Here the ReLu function~\cite{Glorot2011}, defined by $\relu ([z_1,z_2,\cdots,z_{n_e}]^T) =
    [\max\{z_1,0\},\max\{z_2,0\},\cdots,\max\{z_{n_e},0\}]^T$, is a nonlinear function for every neuron. Finally, the neuron(s) in the output layer contains the probabilities for the input state to belong to a specific class. For example, for a binary-state classification, where only one neuron is needed, the output contains the probability for the input state may be identified as entangled or separable.
    
In the following, we shall demonstrate how machine-learning methods can solve the following quantum-state classification problems:
\begin{enumerate}
\item We first consider the question: ``{\it{is it possible to optimize the coefficients $(1,1,1,-2)$ in the CHSH inequality such that the inequality becomes a better state classifier?}}" We found that the answer is positive; in particular, we found that the choice of $(-0.521,0.603,-0.025,0.016,0.373)$ can yield a much better performance for our testing states.

\item Second, instead of linear optimization, we include non-linear elements and a hidden layer. At the same time, we allow the parties to choose random measurement angles. We found that the performance of the state classifier based on the machine-learning method can be enhanced significantly.

\item Third, we ask the question: ``{\it is it possible to construct a universal state classifier for detecting quantum entanglement?}" If possible, this would be a valuable tools for many tasks in quantum information theory. However, the challenge is to find a reliable way for labeling the quantum states in the training set. For a pair of qubits, this is possible by using the PPT (Positive Partial Transpose) criterion. We have constructed such a universal state classifier for a pair of qubits; we found that the performance depends heavily on the training sets; the major source of error comes from the training data at near the boundary between the entangled states and separable states.

\item Next, we consider multiple-state classification involving systems of three qubits. There, the structure is more complicated than two qubits. We constructed a quantum state classifier that can identify four types of separable states.

\item Finally, we considered an ensemble of four-qubit systems. We analyzed the performance of the state classifier in terms of three groups of quantum states. 

\end{enumerate}

\section{Challenges of entanglement detection with CHSH inequality}

\begin{figure}[t]
	\includegraphics[scale=0.3]{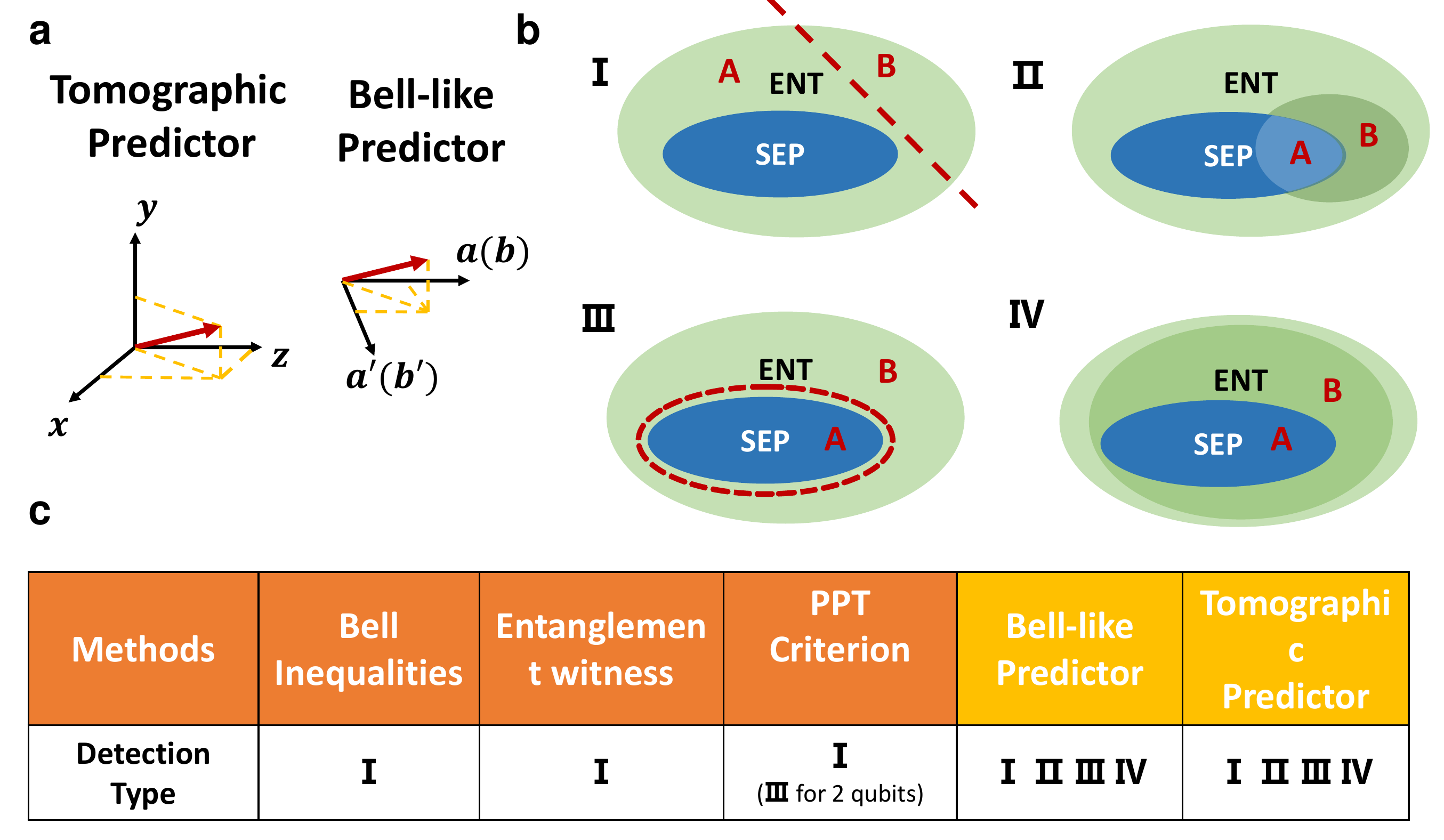}
	\centering
	\caption{\label{fig:Methods}\textbf{Comparison between different methods.}
		\textbf{(a)} Measurement per party.
		For n-qubit system, the process of reconstructing an unknown quantum state by Standard Quantum State Tomography requires 3 measurement~($\sigma_x$, $\sigma_y$, $\sigma_z$) to perform for every qubit, many times each. However, the verification of quantum entanglement by two-setting Bell inequalities only requires 2 operators for each observer. In our work, Tomographic and Bell-like predictors need the same resources to perform as Tomography and Bell inequalities individually.
		\textbf{(b)} State Discrimination.
		Traditional tools to verify entanglement, including Bell inequality, PPT criterion and entanglement witnesses, are capable to identify partial entangled states in multi-qubit system, i.e type type \uppercase\expandafter{\romannumeral 1}. Our methods by machine learning try to identify quantum states from different channel. The method works well if the characteristics of states from different channels on the edge are obviously different.
		\textbf{(c)} Detection type of traditional entanglement detection algorithms (Bell Inq. , PPT, Witness) and our machine learning methods (Bell-like and Tomographic Predictor.)
	}
\end{figure}

\begin{figure}
    \includegraphics[width=0.9\columnwidth]{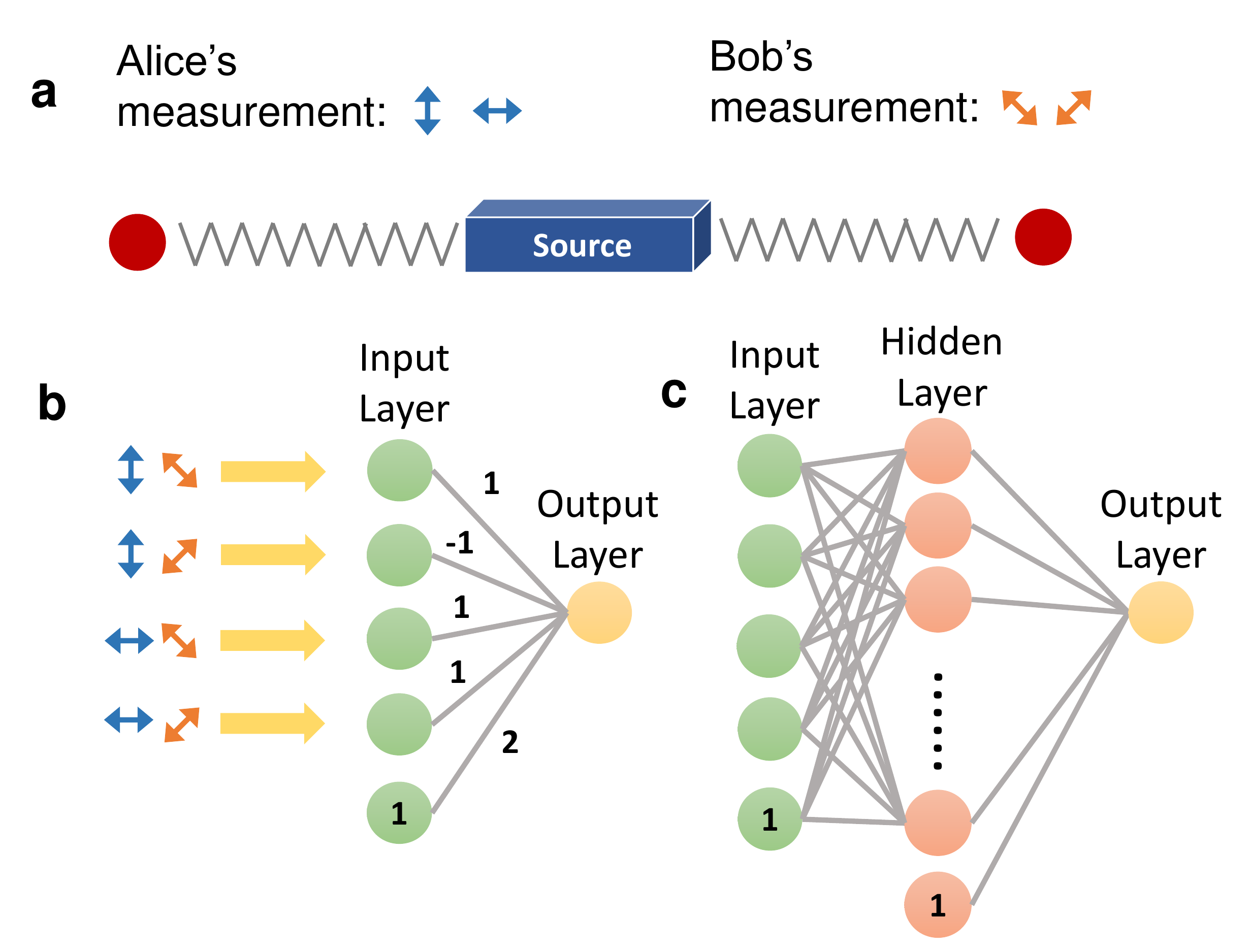}
	\centering
	\caption{\label{fig:Q2_ppt}\textbf{CHSH inequality and machine learning.}
		\textbf{(a)} Typical setup for CHSH inequality.
%
%
		\textbf{(b)} Mapping the CHSH operators to linear artificial neural network (ANN).  
		\textbf{(c)} Artificial Neural Network with Hidden Layers.
		To promote the method to distinguish different quantum states as many as possible, a hidden layer with nonlinear function is inserted between input and output layers. The task is optimizing $\sigma_S(W_2(\relu(W_1\vec{x}+w_{01})) + w_{02})$ where $\relu$ is ReLu~\cite{Glorot2011} function on every hidden neuron {and $\sigma_S(x) = 1/(1 + e^{-x})$ is sigmoid function.}
	}
\end{figure}

\subsection{General background on quantum entanglement}
Entanglement is a key feature of quantum mechanics, where the correlation between pairs or groups of particles cannot be described within a local realistic classical model. In quantum information theory, entanglement is regarded as an important resources to achieving tasks, such as quantum teleportation~\cite{bennett1993teleporting}, quantum computation~\cite{Raussendorf2001}, and quantum cryptography~\cite{Ekert1991}.

However, given a quantum state, the problem of determining if it is entangled or not is a computationally-hard, (NP-hard~\cite{Gurvits2003} more precisely). This question is particularly important in quantum experiments. Currently, methods of entanglement detection has been developed for specific scenarios~\cite{guhne2009entanglement}. The most popular ones includes Positive Partial Transpose (PPT) criteria~\cite{Horodecki1996,Horodecki1997} and entanglement witnesses~\cite{Horodecki1996,Bruss2002,Terhal2000,Lewenstein2000}. 

For a pair of qubits, Positive Partial Transpose (PPT) is both sufficient and necessary for entanglement detection~\cite{banaszek2013focus}. However, PPT is a necessary but not sufficient condition for multi-qubit systems. In addition, it requires the knowledge of the whole density matrix. Experimentally, it means one needs to perform quantum state tomography, which is resource consuming for multiple-qubit systems.

Moreover, entanglement witnesses represent a different approach for entanglement detection. Given an observable $\mathcal{W}$, where $\Tr(\mathcal{W}\rho) \geq 0$ for all separable states. If $\Tr(\mathcal{W}\rho) < 0$ for (at least) one entangled $\rho$, then we say $\mathcal{W}$ detects $\rho$~\cite{guhne2009entanglement,Shahandeh2017}. Here the trace ${\text{Tr}}(\mathcal{W}\rho ) = \left\langle \mathcal{W} \right\rangle$ represents the measurement result of $\rho$ with $\mathcal{W}$. Of course, it is possible that there are entangled states not detected by a given witness, i.e., $\Tr(\mathcal{W}\rho) \geq 0$ for an entangled state.

On the other hand, quantum entanglement is necessary for a violation of Bell's inequalities~\cite{Bell1989}, which has been confirmed in numerous experiments~\cite{Freedman1972,Aspect1981,Ansmann2009,Giustina2013,Hensen2015}. In principle, Bell's inequality can be employed for detecting quantum entanglement; it can witness some entangled states. It is an attractive direction, as only partial information is needed from the quantum state.  However, for normal Bell's inequalities, only small part of the entangled states can be detected; a situation similar to entanglement witness. Motivated by this problem, one of our goals is to construct an quantum-state classifier for entanglement detection through optimizing Bell's inequalities. 

\subsection{Separable states and CHSH inequality}
To get started, let us consider an ensemble of quantum states $\rho$ of $n$ qubits; the method is also applicable for qudit systems. Recall that a quantum state is separable if and only if it can be expressed as a linear combination of product states, i.e.,
\begin{equation}\label{eq:full_seperable_states}
\rho_\text{sep} = \sum_i \ p_i \ \rho_i^1\otimes\rho_i^2\otimes\cdots\otimes \rho_i^n \ .
\end{equation}
Otherwise, the quantum state is entangled. 

In fact, entanglement is necessary for a violation of the Bell inequalities~\cite{Bell1989}, e.g. the CHSH (Clauser-Horne- Shimony-Holt) inequality~\cite{Clauser1969},
\begin{equation}\label{CHSH_original}
\left| \langle \bm{a b} \rangle - \langle \bm{a b'} \rangle + \langle \bm{a' b} \rangle + \langle \bm{a' b'} \rangle \right| \leqslant 2 \ ,
\end{equation}
where $\langle\cdot\rangle$ represents expectation, $\{ \bm{a}, \bm{a'}\}$ and $\{ \bm{b}, \bm{b'}\}$ are the detector settings of parties $A$ and $B$ respectively that take only two values $\pm 1$ (see Fig.~\ref{fig:Q2_ppt}). Furthermore, $\bm{\hat{n}} = n_1 \sigma_x + n_2 \sigma_y + n_3 \sigma_z$, $\bm{\hat{n}} \in \{\bm{a},\bm{a'},\bm{b},\bm{b'}\}$, and $\sigma_{x,y,z}$ are the Pauli matrices. 

Quantum states violating the CHSH inequality can be labeled as ``entangled". However, CHSH inequalities cannot be employed as a reliable tool for entanglement detection. There are two reasons. First, there exist entangled states not violating the Bell inequalities. To be more specific, the maximally-entangled state, such as $\ket{\psi_-} = (\ket{00} - \ket{11}) / \sqrt{2}$ for a pair of qubits, can maximally violate the CHSH inequality~\cite{Bell1989}. However, this tool fails under the circumstances of noise, in the form of a quantum channel. After passing through a depolarizing channel~\cite{nielsen2002quantum}, the resulting state,
\begin{equation}\label{eq:depolarizing_channel}
\rho  = p\left| {{\psi _ - }} \right\rangle \left\langle {{\psi _ - }} \right| + \left( {1 - p} \right)\identity/{4} \ ,  
\end{equation}
where $0 \le p \le 1$, violates the CHSH inequality only if $p > 1/\sqrt{2} \simeq 0.707$~\cite{guhne2009entanglement}. However, the state is entangled when $p>1/3 \simeq 0.333$~\cite{guhne2009entanglement}. 

Another reason is that the measurement angles depends on the quantum state. For example, if we choose fixed measurement angles with the following CHSH operator, 
\begin{equation}\label{CHSH_operator}
\Bellop \equiv \bm{a_0 b_0}-\bm{a_0 b_0'}+\bm{a_0' b_0}+\bm{a_0' b_0'}  \ ,
\end{equation}
where $\bm{a_0}=\sigma_z$, $\bm{a_0'}=\sigma_x$, $\bm{b_0}=(\sigma_x-\sigma_z)/{\sqrt{2}}$, $\bm{b_0}=(\sigma_x+\sigma_z)/{\sqrt{2}}$,
then for any given quantum state of the form, 
\begin{equation}\label{state_thephi_2}
\ket{\DoubleBlochState} = \cos \left( {\theta /2} \right)\left| {00} \right\rangle  + {e^{i\phi }}\sin \left( {\theta /2} \right)\left| {11} \right\rangle \ ,
\end{equation}
we have the expectation value, $\bra{\DoubleBlochState}\Bellop\ket{\DoubleBlochState} = \sqrt{2} \ (\sin\theta\cos\phi - 1)$, which is equal to $-2\sqrt{2}$ when $\theta = \pi/2$, and $\phi = \pi$, i.e., when $\ket{\DoubleBlochState} = \ket{\psi_-}$. For a different value of $\phi$, e.g. $\phi=\pi/2$, the resulting quantum state can no longer be used to violate this particular CHSH inequality. Therefore, in general, the family of (original) CHSH inequalities cannot be employed as a reliable tool for detecting quantum entanglement for given quantum states.

\section{Optimizing CHSH operator with Machine Learning} \label{mlp}

\begin{figure*}
	\includegraphics[scale=0.7]{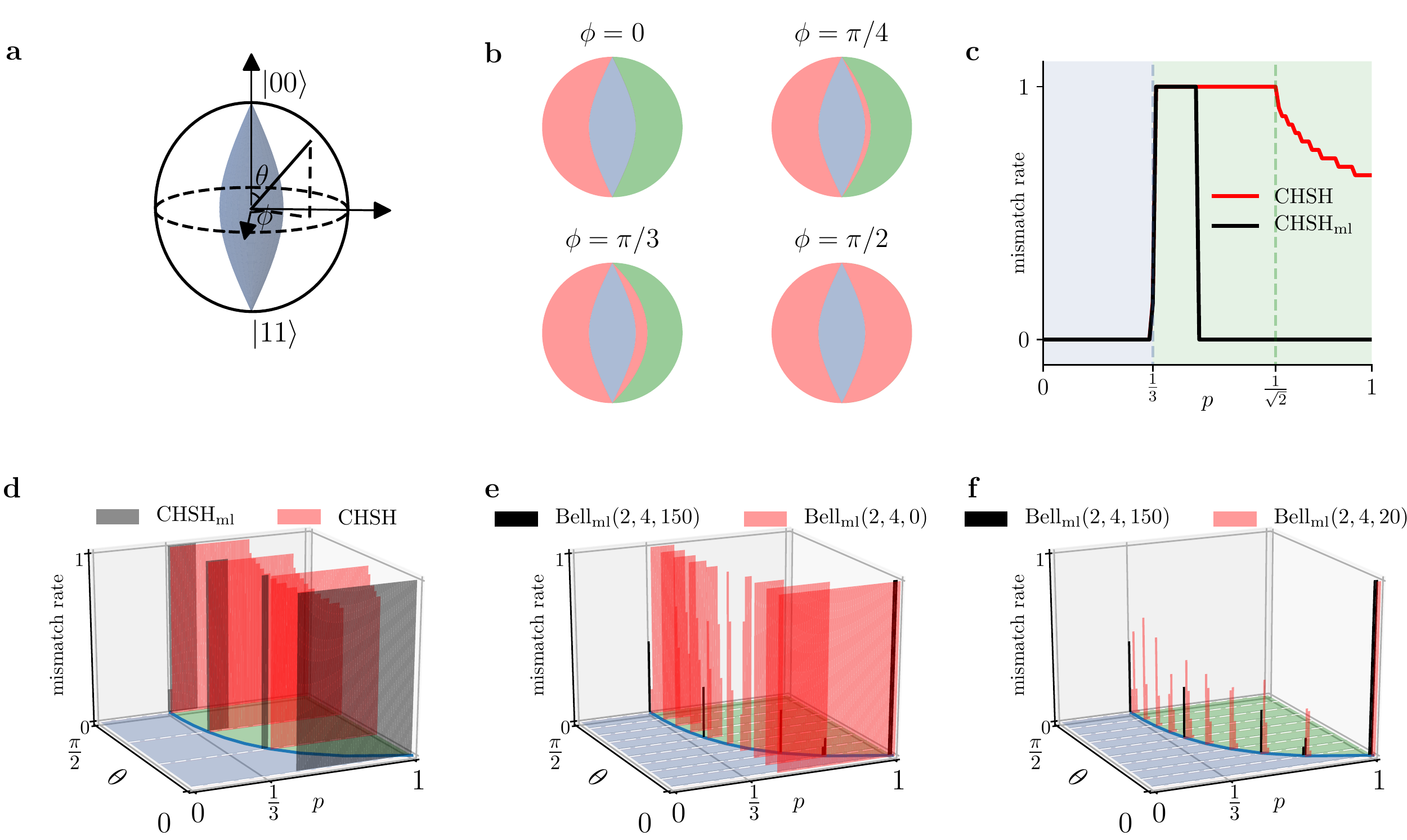}
    \centering
	\caption{\label{fig:Q2_data}\textbf{Results of the first ``test run" by machine learning.}
		\textbf{(a)} The “shape” of quantum states illustrated by Bloch sphere; the blue area represents separable states.
		\textbf{(b)} Limitation of witness detection. The entangled states detected by single witness $\mathcal{W}$ depends on the state phase $\phi$. As an example, the entangled states which lie in green area can be detected, while that in red area can not~(see supplementary).
		\textbf{(c-d)} The optimization of original CHSH inequality by tuning $W$ and $w_0$. 
		For states $\EnsembleForFirst$~(Eq.~(\ref{state_testrun})) with fixed $\theta$, $p$ but different angle $\phi$, the height of vertical axis presents the mismatch rate predicted by PPT criterion and other predictors. Since PPT detects all entangled states in 2-qubit system, the mismatch rate can also be called error rate of entanglement detection. Green and blue areas represent entangled and separable states individually. \textbf{(d)} implies that the new linear predictor~(CHSH$_{\text{ml}}$) intrinsically searches a best critical $p$ to divide entanglement and separable ensembles.
		\textbf{(c)} is the cross section of \textbf{(d)} with $\theta = \pi/2$, which illustrates the optimization of maximum entangled states. 	
		\textbf{(e-f)} Mismatch rate of entanglement detection by Bell-like predictor with different hidden layers~(150,20 or no neurons) on 2-qubit system with random measurement. The mismatch only happens on the edge area of entangled and separable states. For $\theta = 0$, all the ensembles with different $\phi$ degenerate into one state, thus the mismatch rate is either 1 or 0.}
\end{figure*}

In this work, we consider two types of machine learning models to classify different types of quantum ensembles (see Fig.~\ref{fig:Methods}), namely (i) tomographic predictor and (ii)~Bell-like predictor. Tomographic predictors make use of all information of a given quantum state and is used to benchmark the performance of Bell-like predictors, which employs a subset of non-orthogonal measurements setting. For example, for a pair of qubits, the tomographic predictors are the Cartesian product of two sets of Pauli operators, $\{ \identity, \sigma_x, \sigma_y, \sigma_z  \}$, which contains a total of 15 non-trivial combinations. On the other hand, the CHSH operator in Eq.~(\ref{CHSH_operator}) can be regarded as an example of using the Bell-like predictors. 

To elaborate further, we construct a linear Bell-like predictor by generalizing the CHSH operator as (see Eq.~(\ref{CHSH_original}) for notations):  
\begin{equation}\label{CHSH_operator_w}
\Pi_\text{ml} \equiv w_1\bm{a_0 b_0} + w_2\bm{a_0 b_0'} + w_3\bm{a_0'b_0} + w_4\bm{a_0'b_0'} + w_0 \ ,
\end{equation}
where the coefficients (or weights) $\{w_0, w_1, w_2, w_3, w_4 \}$ are determined by the method of machine learning, through minimizing the error of detecting quantum entanglement of a given quantum ensemble. Here the measurement angles $\{\bm{a_0}, \bm{a_0'},\bm{b_0},\bm{b_0'} \}$ are taken to be the same as those given in $\Pi_\text{CHSH}$ defined in Eq.~(\ref{CHSH_operator}). We denote the resulting Bell-like predictor as CHSH$_\text{ml}$. For a given quantum state, the set of observables (called features) 
\begin{equation}
\{ \langle \bm{{a_0}{b_0}} \rangle , \langle \bm{{a_0}{b_0'}} \rangle, \langle \bm{{a_0'}{b_0}} \rangle, \langle \bm{{a_0'}{b_0'}} \rangle  \} \ ,
\end{equation}
are taken as the input of the machine learning program; normally, the number of elements in this set should be much smaller than the dimension of the quantum state. 

In fact, the method of machine learning allows us to construct more general Bell-like predictors, given the same number of features. The key element of them is the inclusion of an extra hidden layer of neurons (see Fig.~\ref{fig:Q2_ppt}), compared with the linear predictor CHSH$_\text{ml}$. Moreover, each link between a pair of neurons is associated with a weight to be optimized in the learning phase.

Specifically, here we consider a class of (non-linear) predictors denoted by 
\begin{equation}
\text{BELL}_\text{ml} (n,n_f,n_e),
\end{equation}
where $n$ labels the number of qubits in the quantum state, $n_f$ labels the number of features, and $n_e$ labels the number of neurons in the hidden layer of the neuron network. Apart from the extra neurons in the hidden layer, the measurement angles $\{ \bm{a},\bm{a'},\bm{b},\bm{b}' \}$ in the corresponding feature list are taken randomly.

As the first ``test run" of our machine learning method, we focus on the following family of quantum states:
\begin{equation}\label{state_testrun}
\EnsembleForFirst = p\ket{\DoubleBlochState}\bra{\DoubleBlochState} + (1-p) I/4  \ ,
\end{equation}
where $\ket{\DoubleBlochState}$ is defined in Eq.~(\ref{state_thephi_2}), $0 \le p \le 1$. For a pair of qubits, the entanglement between them can be determined by checking the PPT (positive partial transpose) criterion~\cite{Horodecki1996,Horodecki1997}: let ${\rho _{\theta ,\phi }^{{T_B}}}$ be the matrix obtained by taking partial transpose of $\rho_{\theta,\phi}$ {in the second qubit}. The state is entangled if and only if the smallest eigenvalue of the matrix ${\rho _{\theta ,\phi }^{{T_B}}}$ is negative.

For our case, the minimal eigenvalue can be obtained analytically (see supplementary materials), which is given by
\begin{equation}\label{min_eigva}
{\lambda _{\min }} ( {\rho _{\theta ,\phi }^{{T_B}}}) = \left( {1 - p} \right)/4 - p\cos \left( {\theta /2} \right)\sin \left( {\theta /2} \right) \ .
\end{equation}
For each quantum state in the training set, we first evaluate the value of ${\lambda _{\min }} ( {\rho _{\theta ,\phi }^{{T_B}}})$, in order to create a label for it. In Fig.~\ref{fig:Q2_data}\textbf{a}, we depict the portion of separable states in the colored area of a Bloch sphere.

\subsection{Training phase of the predictors}

To investigate the performance of CHSH$_\text{ml}$, which is essentially a linearly-optimized version of CHSH, and a non-linear predictor with machine learning (see Fig.~\ref{fig:Q2_ppt}\textbf{b}). First, we need to generate an initial set of quantum states, called training set. {The set of states are generated by sampling a uniform distribution of $\theta$ and $\phi$, but with a Gaussian distribution for $p$, with a mean value $1/(1+2\sin\theta)$, which yields an ensemble of states in the neighborhood of separable and entangled hyperplane.

Specifically, for each time, we evaluated the four features,
$\{ \langle { \bm{a_0 b_0}} \rangle, \langle { \bm{a_0 b_0'}} \rangle , \langle { \bm{a_0' b_0}} \rangle , \langle { \bm{a_0' b_0'}} \rangle  \} $, in the CHSH$_\text{ml}$ for a given state in the training set, putting them into a four-dimensional feature vector $\vec{x}$ in ANN (artificial neural network). In fact, if we consider only one side of the inequality, the CHSH inequality is equivalent to 
\begin{equation}
W_0 \ \vec{x} + w_0 \ge 0 \ ,
\end{equation}
where $W_0 = [1,-1,1,1]$ and $w_0 = 2$.
In other words, CHSH inequality are violated iff the output value is negative. The problem of optimization of CHSH$_\text{ml}$ is equivalent to the problem of finding an optimal set of matrix elements for $W$ and $w_0$, through the given training set of quantum state.

The training steps are as follows: first, {we apply a sigmoid function,
\begin{equation}
\sigma_S(x) = 1/(1 + e^{-x}) \ ,
\end{equation}
for the output layer (Fig.~\ref{fig:Q2_ppt}\textbf{(b,c)}). The output value  represents the separable set.} Then, we make use of a loss function constructed by the cross entropy~\cite{dunne1997pairing} to calculate the difference between predictor and the results based on the PPT criterion for many copies in the given quantum ensemble. Next, the loss function was minimized using the stochastic gradient descent algorithm~\cite{bottou2010large}. At the end, we obtained a vector $W$ and $w_0$ that is optimized by the above process. 

\begin{table*}[t!]\label{tb} 
	\caption{CHSH inequality versus machine learning predictors for two qubits }  	
    \begin{tabular*}{14cm}{llll}  
	\hline  
	\textbf{Type} & \textbf{Features}  & \textbf{weights} & \textbf{form} \\
	\hline   
	CHSH  & fixed $\{\bm{a_0 b_0}, \bm{a_0 b_0'}, \bm{a_0' b_0}, \bm{a_0' b_0'}\}$\footnote{$\bm{a_0}=\sigma_z$, $\bm{a_0'}=\sigma_x$, $\bm{b_0}=(\sigma_x-\sigma_z)/{\sqrt{2}}$, $\bm{b_0}=(\sigma_x+\sigma_z)/{\sqrt{2}}$} &  4 fixed values\footnote{$w_0 = \pm 2, w_1 = 1, w_2 = -1, w_3 = 1, w_4 = 1$ } & linear, no hidden layer \\  
	\CHSHML  & fixed $\{\bm{a_0 b_0}, \bm{a_0 b_0'}, \bm{a_0' b_0}, \bm{a_0' b_0'}\}$ & 4 optimized values & linear, no hidden layer \\
	BELL$_\text{ml}$ & variable $\{\bm{a b}, \bm{a b'}, \bm{a' b}, \bm{a' b'}\}$ \footnote{$\bm{a}$, $\bm{b}$, $\bm{a'}$, and $\bm{b'}$ are randomly generated} & many optimized values & non-linear, with hidden layer \\
	\hline
\end{tabular*}
\end{table*}

\subsection{Testing phase of the predictors}

After the {predictor} is well-trained, we test the performance by creating a new set of quantum ensemble that is distinct from the {data} set employed for training. Here the testing data comes from an ensemble of quantum states $\EnsembleForFirst$ with a uniform distribution of $p$, $\theta$ and $\phi$. Note that from Eq.~(\ref{min_eigva}), the entanglement of $\EnsembleForFirst$ depends on the values of  $p$ and $\theta$ but not $\phi$. However, the same set of features of the new density matrices are provided as the input; the values of $p$ and $\theta$ are not directly provided  in the testing phase, but they are used to evaluate the performance of the predictors.

We quantify the performance of the CHSH$_\text{ml}$ predictor as follows: for given values of $p$ and $\theta$, the mismatch rate ${R_{{\text{mm}}}}\left( {p,\theta } \right)$ is defined by the probability that the function outputs a different label from the PPT criterion, averaged over uniform distribution of the angle $\phi$, i.e.,
\begin{equation}
{R_{{\text{mm}}}}\left( {p,\theta } \right) \equiv \Pr \left( {{1_{{\text{ML}}}}|{0_{{\text{PPT}}}}} \right) + \Pr \left( {{0_{{\text{ML}}}}|{1_{{\text{PPT}}}}} \right) \ ,
\end{equation}
where ${x_{{\text{ML}}}} \in \{ 0 ,1 \}$ labels the output of the machine learning predictor; ${{1_{{\text{ML}}}}}$ (${{0_{{\text{ML}}}}}$) means separable (entangled), and similarly for ${x_{{\text{PPT}}}}$. Of course, the match rate can be defined in a similar way.

The {mismatch} rates for the predictor CHSH$_\text{ml}$ are shown in Fig.~\ref{fig:Q2_data}\textbf{(c,d)}. We also include the use of CHSH inequality for entanglement detection for a comparison. 
The numerical data indicates that both CHSH and CHSH$_\text{ml}$ can identify the regime where ${\lambda _{\min }} ( {\rho _{\theta ,\phi }^{{T_B}}}) > 0$ as separable, except when $\theta$ is close to zero $\theta=0$. { When $\theta=0$, the state $\ket{\DoubleBlochState}$ reduces to only one state $\ket{00}$ for any choice of $\phi$. Therefore, the mismatch rate becomes $100\%$ whenever the predictor made a mistake.} We shall see that such a problem in CHSH$_\text{ml}$ also exists in the BELL$_\text{ml}$ predictor without hidden layer. However, the problem goes away whenever hidden layers are included.

Beyond that region, CHSH produces $100\%$ mismatch rate, but CHSH$_\text{ml}$ can significantly reduce the mismatch rate as $p$ increases. The reason for the CHSH to produce $100\%$ mismatch rate is similar to the situation explained after Eq.~(\ref{eq:depolarizing_channel}); there exist entangled states not violating the CHSH inequality. 

Therefore, we conclude that performance of CHSH$_\text{ml}$ is significantly better than that of CHSH in detecting quantum entanglement in the regime ${\lambda _{\min }} ( {\rho _{\theta ,\phi }^{{T_B}}}) < 0$. In CHSH$_\text{ml}$, the measurement angle is fixed, we shall see that the performance of machine learning can be significantly increased if we choose to make the measurement angle random. 

The result of predictor, BELL$_\text{ml} (2,4,0)$ (i.e., for 2 qbuits, 4 features, and 0 hidden layer) is shown in Fig.~\ref{fig:Q2_data}\textbf{(e,f)}. We have also included the results of BELL$_\text{ml} (2,4,20)$ and BELL$_\text{ml} (2,4,150)$ predictors, with respectively 20 and 150 neurons in the hidden layer, for comparison. The overall performance in terms of mismatch rates are significantly improved, compared with the CHSH$_\text{ml}$ predictor. Furthermore, inclusion of a hidden layer can significantly mitigate the problem of CHSH$_\text{ml}$ near $\theta = 0$. Numerically, we found that the results with a total of 150 neurons in the hidden layer do not significantly out perform the results with 20 neurons.


\section{Classifying general two-qubit states}
\label{sec:gap_data}

\begin{figure}[t]
	\includegraphics[scale=0.5]{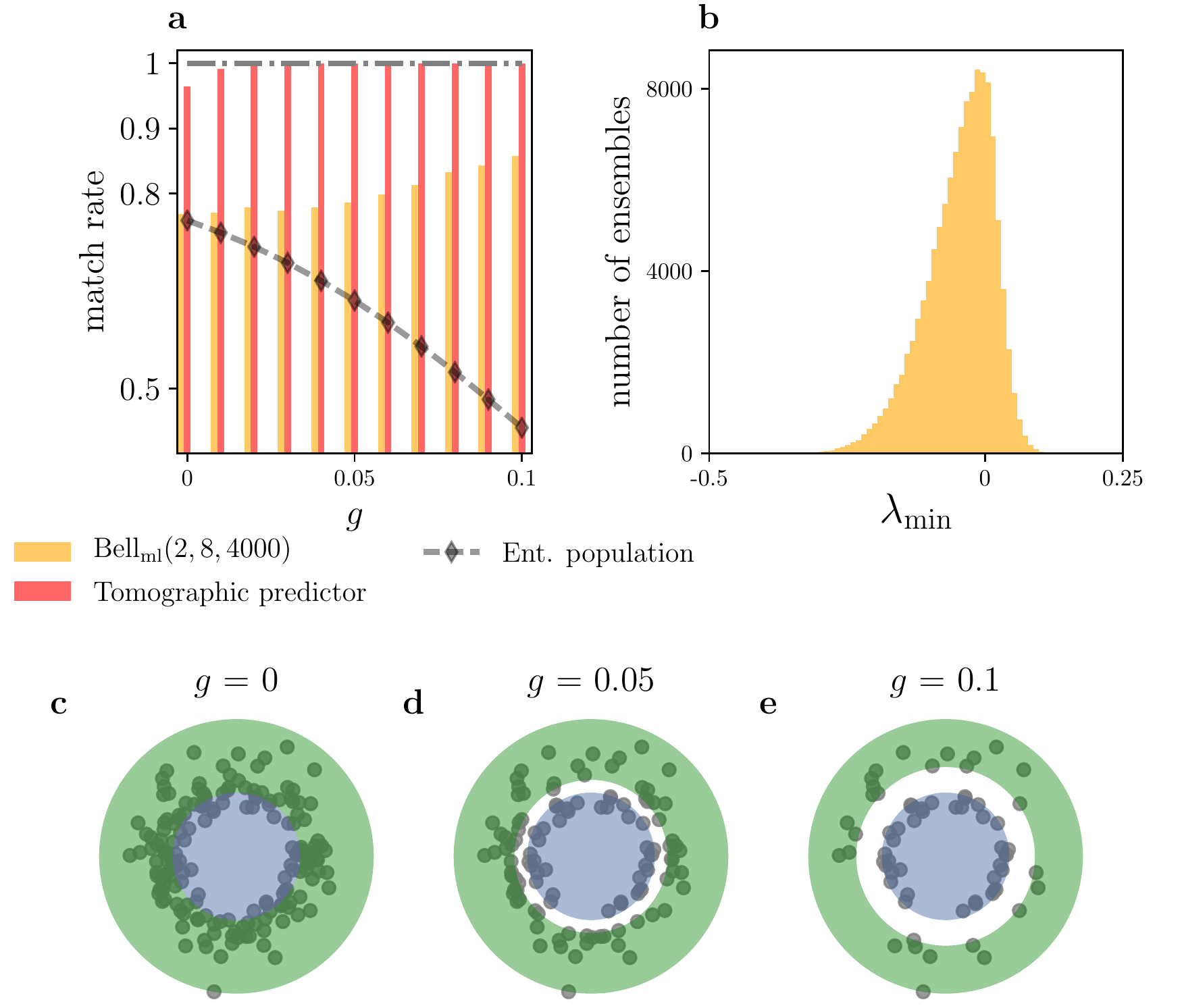}
	\centering
	\caption{\label{fig:Gap_data}\textbf{Bell-like $\&$ Tomographic predictor on all 2-qubit states.}
		\textbf{(a)} Match rate of machine learning predictors on the 2-qubit ensembles that generated according to Harr distribution and then filtered out the entangled states that close to separable ones. The ratio of entanglement ensembles to all states falls below $0.5$ with $g>0.1$. The match rate of Tomographic predictor grows up to $99.9\%$ when $g>0.01$, and that of Bell-like predictor also grows with $g$ increasing.   
		\textbf{(b)} Distribution of the minimal eigenvalue $\lambda_\text{min}(\rho_{AB}^{T_B})$ of the matrix $\rho_{AB}$ after a partial transpose.
        \textbf{(c-e)} illustrate the state distribution. Green area is for entangled states; blue area is for separable states. 
	}
\end{figure}

In the previous section, we have studied the ability of the machine-learning predictors in detecting the entanglement of quantum states of the form indicated in Eq.~(\ref{state_testrun}), which belongs to the type \uppercase\expandafter{\romannumeral 2} problem. A potentially more interesting question is, can we construct a universal function that accepts only partial information about the quantum state but, at the same time, can detect all entangled states by machine learning (i.e., type \uppercase\expandafter{\romannumeral 3})? For the case of two qubits, we can still rely on the PPT criterion to provide labels for our training set. 

For this part, we generate a new training set of 2-qubit mixed states randomly and label them by the PPT criterion in the same way as the previous section. The ensembles $\rho$ are prepared by first generating a set of random matrices $\sigma$, where the real and imaginary parts of the elements $\sigma_{ij} = a_{ij} + i b_{ij}$ are generated by a Gaussian distribution with a zero mean and unit variance. The resulting density matrix is obtained by
\begin{equation}
{\rho _{{\text{rand}}}} = \sigma {\sigma ^\dag }/{\text{tr}}(\sigma {\sigma ^\dag }) \ ，
\end{equation}
which is implemented by using the code of QETLAB~\cite{qetlab}.

The performance of our machine-learning predictors heavily depends not only on the training set but also the distribution of the testing data. We found that many data points (of quantum states) are localized near the boundary between entangled and separable states, which represents a challenge for us; machine learning performs not so well near the marginal cases. To regulate the anomaly, we introduce a gap in the boundary between the entangled and separable states. 

Here the gap is defined for the entangled states where the absolute value of the minimum eigenvalue $\lambda_\text{min} (\rho_{AB}^{T_B})$ of the partial transposed matrix  $\rho_{AB}^{T_B}$ is larger than a given constant $g$, i.e.,
\begin{equation}
\left| {{\lambda _{\min }}} \right| > g \ .
\end{equation}
The distribution of $\lambda_\text{min}$ in our data set is given in Fig.~\ref{fig:Gap_data}\textbf{b}. We can see that the majority of states are weakly entangled, which imposes the challenge for our machine-learning predictors.

For this setting, we first present the results of the Bell$_\text{ml}(2,8,4000)$ predictor containing 4000 neurons in the hidden layer. As in shown in (Fig.~\ref{fig:Gap_data}\textbf{a}), the match rate is about 75\% when a gap is not introduced in the testing data. Unfortunately, the match rate cannot be considered as high, as the population of entangled states in the testing set is also about 75\%; one can achieve the same performance by guessing all given states as entangled. This implies that the number of features involved in Bell$_\text{ml}(2,8,4000)$ are not sufficient to detect entangled states in the ensemble. For a comparison, we consider using the Tomographic Predictor, which takes all $15$ combinations of the two-qubit Pauli operators as an input. The result of match rate goes up to larger than $96\%$. This result implies that machine learning becomes more reliable when more information about the quantum states are available. 

In addition, our Bell-like predictor gets improved if the entangled states near the boundary of the separable states are filtered out, as shown in Fig.~\ref{fig:Gap_data}\textbf{a}. Here we vary the gap from 0 to 0.1. The larger the gap, the better the performance of the Bell-like predictor. When the gap is about 0.07, the fraction of the entangled states is about the same as separable states, the match rate of the Bell-like predictor is about 80\%.

\section{Three-qubit systems} \label{multi-qubit}

\begin{figure}[b]
	\includegraphics[width=0.9\columnwidth]{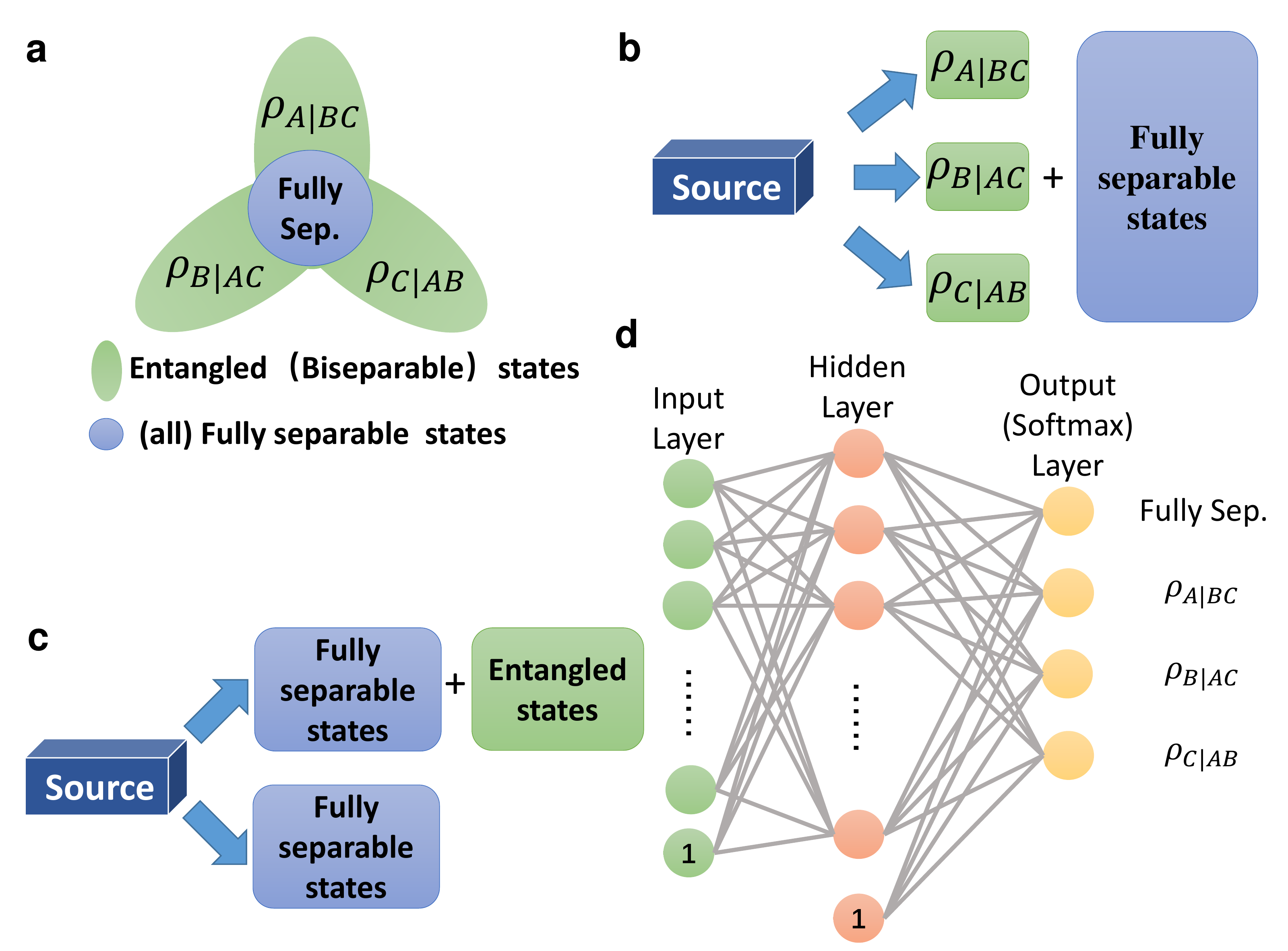}
	\centering
	\caption{\label{fig:Q3Q4_ppt}\textbf{Multi(3 and 4)-qubit system.} 
		\textbf{(a)} The relationship between 3 types of biseparable quantum states and fully separable states. 
		\textbf{(b)} The generation process of 3-qubit biseparable states.
		\textbf{(c)} The generation process of 4-qubit system.
		One channel is assumed to generate random fully separable states, while the other generate random entangled particles but mixed with separable states (noise).
		\textbf{(d)} The ANN to distinguish multi-class quantum states. We apply the model to detect the entanglement of 3-qubit system.
		The output are Softmax layer $[p_0, p_1, p_2, p_3]^T(\sum_{i=0}^3 p_i=1)$ which describes the probability of every possible state.
	}
\end{figure}

\begin{figure}
    \includegraphics[width=0.99\columnwidth]{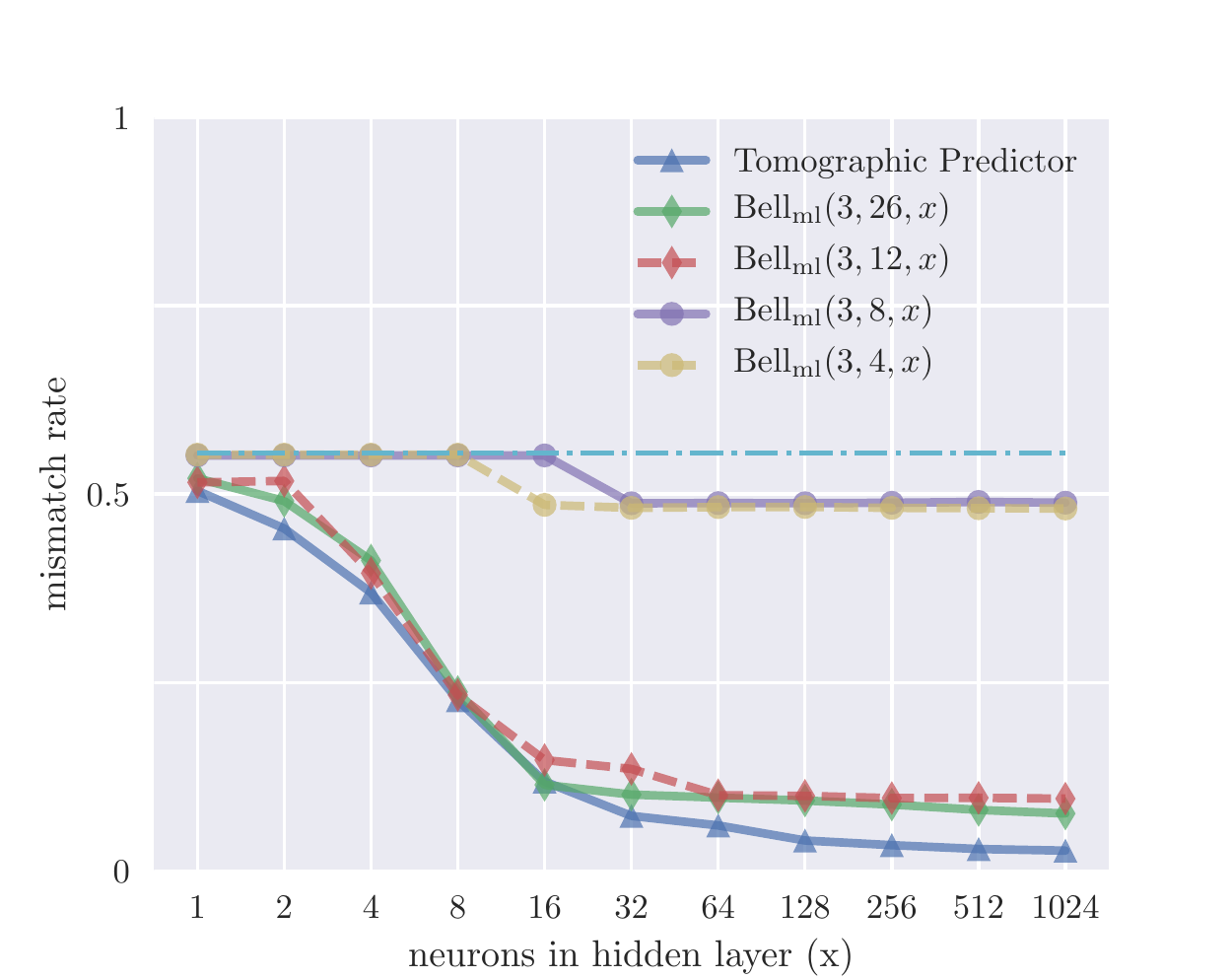}
	\centering
	\caption{\label{fig:Bisep_data}\textbf{Biseparable states distinguished via Bell-like predictors.}
		5 types of predictors applied on the biseparable ensembles generated according to Fig. \ref{fig:Q3Q4_ppt}\textbf{a}. 
        The mismatch rate of our predictor decreases with the number of hidden layer neurons growing. The predictors whose features are equivalent to the Bell inequalities that only detect genuine entangled states, such as Bell$_{\text{ml}}(3,4,x)$ and Bell$_{\text{ml}}(3,8,x)$, almost fail to distinguish biseparate and fully separate states in our problem. On the contrary, the predictors with 3 groups of CHSH features, such as Bell$_{\text{ml}}(3,12,x)$, perform well.
	}
\end{figure}

In general, when extending our method of machine learning to multiple-qubit quantum states, the challenge is to find an efficient way to label the qubits.  


The entanglement structure of a three-qubit system is significantly more complicated than two-qubit systems; it can be classified into several types of entanglement classes~\cite{guhne2009entanglement}. In particular, a three-qubit quantum state is called ``biseparable",  if two of the qubits are entangled with each other but not with the third one. The corresponding density matrices are denoted as,
\begin{equation}\label{rhoABCACBBCA}
\{ \  \rho_{A|BC}, \quad \rho_{B|AC}, \quad \rho_{C|AB} \ \} \ ,
\end{equation}
and their convex combination, i.e.,  $p_1\rho_{A|BC}+p_2\rho_{B|AC}+p_3\rho_{C|AB}$ for $0 \le p_1,p_2,p_3 \le 1$. Of course, these sets of states include fully-separable states as a special case. A system is called fully entangled (or genuine tripartite entangled for pure states)~\cite{guhne2009entanglement} if it is neither biseparable nor fully separable. 

In the following, we shall focus on the quantum ensemble of the following form (see Fig.~\ref{fig:Q3Q4_ppt}\textbf{(a,b)} ),
\begin{equation}\label{eq:biseparable_states}
\rho = p\ket{\psi_{bs}}\bra{\psi_{bs}} + (1-p)\rho_\text{sep} \ ,
\end{equation}
where $\ket{\psi_{bs}}$ belongs to a random pure state of one of the three biseparable states in Eq.~(\ref{rhoABCACBBCA}), the value of $p \in [0,1]$ is generated uniformly, and $\rho_\text{sep}$ is a random fully separable states defined in Eq.~(\ref{eq:full_seperable_states}). The set of states $\ket{\psi_{bs}}$ is generated by random vectors as follows. A random vector is defined by a vector,
\begin{equation}\label{random_vectors_v123}
{\bf v}_\text{rand}(d) = \{ v_1, v_2, v_3, ...\} \ ,
\end{equation}
containing $d$ elements $v_i$ from the Gaussian distribution with a zero mean and unit variance.  For example, for the case of $\rho_{A|BC}$, the state $\ket{\psi_{bs}}$ is obtained by the tensor product ${\bf v}_\text{rand}(2) \otimes {\bf v}_\text{rand}(4)$, followed by multiplying a normalization factor.

In this case, we can still label the entanglement of the quantum states by the use of the PPT criterion. For example, if the pure state $\ket{\psi_{bs}}$ is in $\rho_{A|BC}$, then we can trace out the system $A$ in the total density matrix of the form given in Eq.~(\ref{eq:biseparable_states}), then apply the PPT criterion to the reduced density matrix of $BC$. In this way, we challenge our Bell-like predictors to classify four types of states, including 3 types of biseparable entangled states and the fully-separable states.

In our implementation, similar to our previous construction of the Bell-like predictors based on the Bell's inequalities. Here we consider the Mermin inequality~\cite{mermin1990extreme} and  Svetlichny inequality~\cite{svetlichny1987distinguishing}. {For three-qubit systems, the Mermin inequality is of the form, 
\begin{equation}
\left| \langle \bm{abc} \rangle -\langle \bm{a'b'c} \rangle - \langle \bm{ab'c'} \rangle - \langle \bm{a'bc'} \rangle \right| \le 2 \ ,
\end{equation}
The Svetlichny inequality (essentially a double Mermin inequality) is of the following form: 
\begin{eqnarray}
& & | \langle \bm{abc'} \rangle + \langle \bm{ab'c} \rangle + \langle \bm{a'bc} \rangle - \langle \bm{a'b'c'} \rangle + 
\nonumber \\ 
&& \langle \bm{a'b'c} \rangle + \langle \bm{a'bc'} \rangle +\langle \bm{ab'c'} \rangle - \langle \bm{abc} \rangle | \le 2
\end{eqnarray}
The Mermin inequality and the Svetlichny inequality are the multipartite counterpart of Bell's inequalities. Therefore, one can also employ these inequalities for detecting multipartite entanglement. In a similar way, we can also apply the machine learning method to boost the efficiency.

In our machine learning method, we adopted the elements of Mermin inequality in the photonic experiments of Refs.~\cite{pan2000experimental} as input (4 features) to train our Bell-like predictor Bell$_\text{ml}(3,4,x)$, and similarly, for the Svetlichny inequality, we constructed Bell$_\text{ml}(3,8,x)$). In the cases discussed previously, the sigmoid function was employed for binary classification of the quantum states. Here the output layer is obtained by the Softmax function~\cite{dunne1997pairing}, which can be applied to multi-state classification (Fig.~\ref{fig:Q3Q4_ppt}\textbf{d}).

The mismatch rate of the machine learning method is shown in Fig.~\ref{fig:Bisep_data}. It is indicated that if we just use the same number of features as in Mermin Bell$_\text{ml}(3,4,x)$ and Svetlichny Bell$_\text{ml}(3,8,x)$ inequalities, the performance is not satisfactory. The mismatch rate cannot get much improved by increasing the number of neurons in the hidden layer. However, the performance of the machine learning method can get significantly improved by putting 3 groups of features from CHSH inequalities for every pair of qubits, which gives a new Bell$_\text{ml}(3,12,x)$ predictor. 

{Furthermore, We also trained our model, with both Tomographic predictor and Bell$_{\text{ml}}(3,26,x)$, to distinguish separable states from entangled states which cannot be identified by the PPT criterion (i.e. bound entangled states), such as the entangled states generated by unextendible product basis (UPB)~\cite{bennett1999unextendible}. The accuracy on the test data is larger than $99\%$.}

\section{Four-qubit systems}
\begin{figure}[t]
	\includegraphics[scale=0.7]{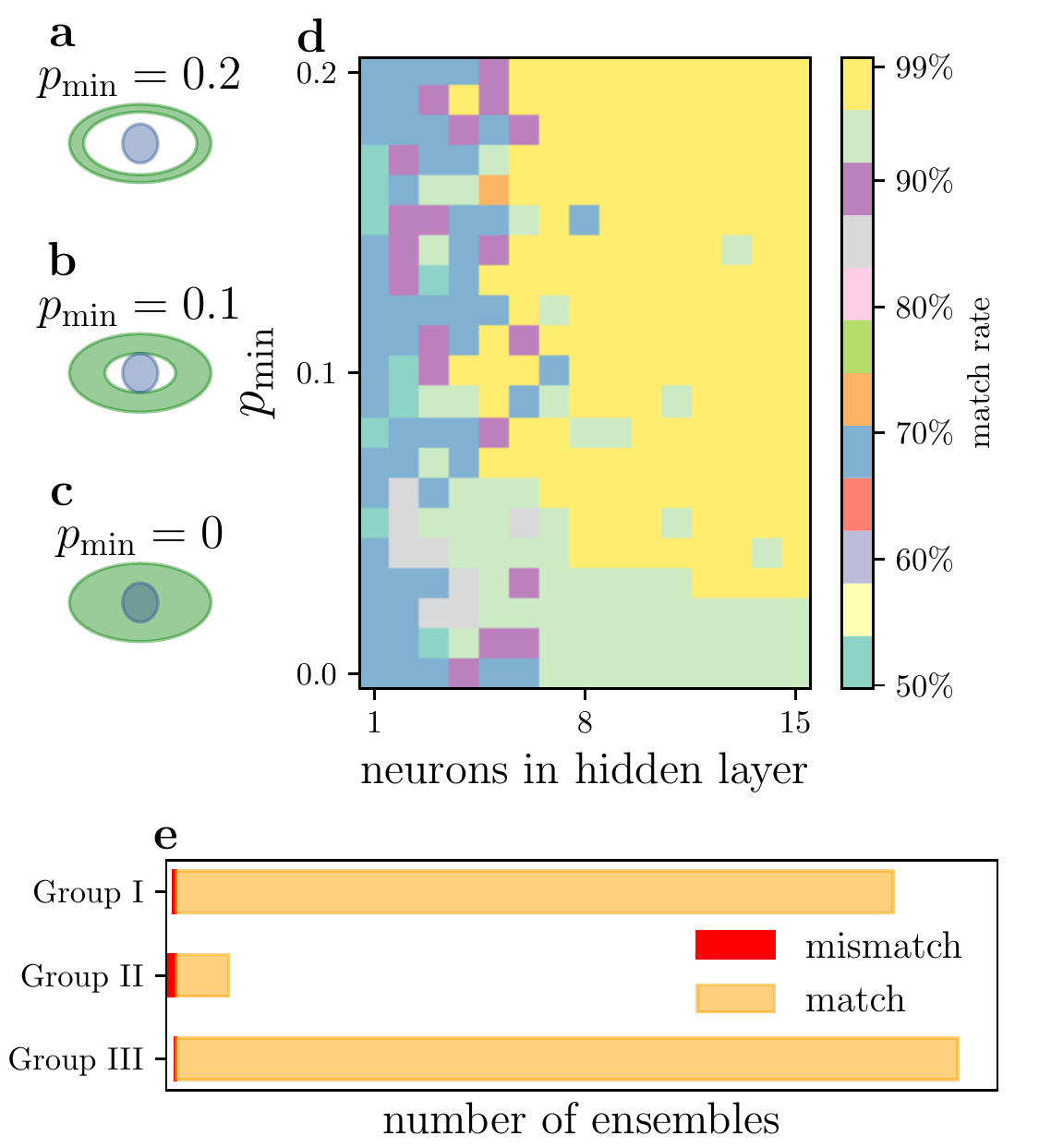}
	\centering
	\caption{\label{fig:Q4_data}\textbf{Quantum states from two channels distinguished by Bell-like(\textbf{d}) and Tomographic(\textbf{e}) predictors.}
		\textbf{(a-c)} The diagram of states from two different channels (Fig.~\ref{fig:Q3Q4_ppt}). Blue channel represents fully separable states and green channel are entangled states with $p$ larger than $p_{\min}$.
		\textbf{(d)} The match rate of states classification by Bell$_\text{ml}(4,80,x)$ with $p_{\min}$ ($y$-axis) and number of hidden layer neurons ($x$-axis) increasing.
		\textbf{(e)} The match rate of prediction by Tomographic predictor and PPT criterion when $p_{\min} = 0$.
	}
\end{figure}

Finally, we apply our machine learning method to the case of a four-qubit system to classify two special classes of quantum states (Fig.~\ref{fig:Q3Q4_ppt}\textbf{c}). Ensembles from blue channel is a fully-separable state $\rho_\text{sep}$ (see Eq.~(\ref{eq:full_seperable_states})) and the others in green is assumed to be the form
\begin{equation}\label{rho_mix_green_p}
\rho_\text{mix} = p\ket{\psi_\text{rand}}\bra{\psi_\text{rand}} + (1-p)\rho_\text{sep} \ ,
\end{equation}
where $\ket{\psi_\text{rand}}$ is a totally random pure state (defined in Eq.~(\ref{random_vectors_v123})) and $p \in [p_{\min}, 1]$ is a uniformly random variable. Note that we need to set a minimum value for $p_{\min}$; it is because if $p_{\min} = 0$, then $\rho = \rho_\text{sep}$, making the two sets of states identical.

Recall that the PPT criterion identifies entangled state whenever the eigenvalue of the corresponding matrix, after partial transpose, becomes negative. Here we found numerically that when $p_{\min} > 0.1$, all of the instances in $\rho_\text{mix}$ are entangled states. The scenarios is depicted in Fig.~\ref{fig:Q4_data}\textbf{(a-c)}. 

As shown in Fig.~\ref{fig:Q4_data}\textbf{d}, we studied the performance of a class of Bell-like predictors, namely {Bell$_\text{ml}(4, 80, x)$, where the neuron number in hidden layer is taken from 1 to 15, i.e., $x = 1,2,\cdots 15$. The 80 features are generated by the following way: assume there are four parties and each party performs a measurement on a qubit locally in two different angles labeled by $\bm{\hat{n}^{i}}$, $\bm{\hat{n}'^{ i}}$ ($i = 1,2,3,4$). Then a feature is obtained by the joint expectation value $\langle \bm{O^1O^2O^3O^4}\rangle$, where $\bm{O} \in \{\bm{\hat{n}}, \bm{\hat{n}'}, \identity \}$. Note that the special case  $\identity^1\identity^2\identity^3\identity^4$ is excluded, since $\langle \identity^1\identity^2\identity^3\identity^4 \rangle = 1$ for any quantum state.} 

When the number of neurons in the hidden layer becomes sufficiently  large, the Bell-like predictor is capable of distinguishing the two classes of states with more than $99\%$ in match rate, when $p_{\min} = 0.1$. In other words, the ensembles assumed to be separable or entangled can be classified reliably by our predictor with an accuracy higher than $99\%$. The match rate can reach $95\%$ even if $p_{\min}=0$, where some of the states in $\rho_\text{mix}$ becomes separable. 

To investigate further, we divide the data into three groups.
\begin{itemize}
\item {\bf Group I:} The subclass of quantum states in $\rho_\text{mix}$ in Eq.~(\ref{rho_mix_green_p}), in which the minimal eigenvalue, of the partial-transposed density matrix, is negative. These states are all entangled.
\item {\bf Group II:} The complementary class of states of group I for $\rho_\text{mix}$, i.e., with non-negative eigenvalues. These states should contain both entangled and separable states. 
\item {\bf Group III:} The class of all fully-separable quantum states~ $\rho_\text{sep}$ (see Eq.~(\ref{eq:full_seperable_states})). These states are all separable.
\end{itemize}

Our goal is to study the performance of the machine learning method in analyzing the internal structure of mixed quantum states. In the training phase, we labeled all states in group I and II with the label {\sf Green}, and states in group III with {\sf Blue}. Fig.~\ref{fig:Q4_data}\textbf{e} illustrates the result of the states with $p_{\min}=0$, trained by the Tomographic predictor. The length of bars represents the number of ensembles tested. For group I, which are definitely entangled, our Tomographic predictor detects their entanglement with more than $99.6\%$ (Ent. in figure) accuracy in terms of the match rate, which grows up to $99.9\%$ (Sep. in figure) for the fully-separable states in group III. For group II, although the states were labeled as {\sf Green} (which is dominated by entangled states), Tomographic predictor suggests that a large fraction of the states are actually {\sf{Blue}} (separable) states, which is consistent with the PPT criterion. 


\section{Conclusion}
In this work, we have applied a method of machine learning, known as Artificial Neuron Networks (ANN), to solve problems of quantum state classification in quantum information science. We have achieved several results, including (i) linear optimizing CHSH inequality, (ii)  nonlinear optimization of the Bell-type inequalities, (iii) construction of universal entanglement detector for two-qubit systems, (iii) multiple-state classification for three-qubit systems, and (iv)  four-qubit systems. Overall, we found that machine-learning can produce reliable results provided that the training set is properly chosen. The performance of machine learning becomes worse whenever the majority of the quantum states in the training set lies around the boundary between two classes (e.g. entangled and separable) of quantum states. 

In general, our results are useful for problems where the process of labeling a quantum state is resource consuming. For example, the use of the PPT criterion requires a diagonalization of an exponentially-large matrix for $n$ qubits. However, these procedures can be confined to the problems in the training set. In the future, we can imagine that these resource-demanding tasks can be achieved by a few super (quantum or classical) computers. Once a predictor is constructed, any small laboratory can makes use of it by measuring only a relatively small amount of features of a testing quantum states. In this sense, the task of solving the computational problems can be shared by users with different computational powers.

\end{document}


\preprint{APS/123-QED}

\title{Supplementary Information ``Transforming Bell's Inequalities into State Classifiers with Machine Learning"}%

\author{Yue-Chi Ma}
 \affiliation{Center for Quantum Information, Institute for Interdisciplinary Information Sciences, Tsinghua University, Beijing,100084, China}
\author{Man-Hong Yung}%
 \email{yung@sustc.edu.cn}
 \affiliation{Department of Physics, Southern University of Science and Technology, ShenZhen 518055, China}
 \affiliation{Center for Quantum Information, Institute for Interdisciplinary Information Sciences, Tsinghua University, Beijing,100084, China}


\maketitle

\tableofcontents

\newpage

\section{\label{sec:PPT} Positive Partial Transpose (PPT)}
In this section, we will introduce Positive Partial Transpose (PPT) criterion~\cite{Horodecki1996,Horodecki1997}, which is an analytic algorithm to ensure an ensemble $\rho$ to be entangled. As an example, we then apply PPT to a specific ensemble, which we used to demonstrate our machine learning method in the main text.

We can expand any ensemble $\rho$ of 2 quantum systems $A$ and $B$ in a given product basis as
\begin{eqnarray}
\rho = \sum_{i,j,k,l}\rho_{kl}^{ij}\ket{i}\bra{j}\otimes\ket{k}\bra{l}
\end{eqnarray}
The partial transposition (with respect to the second party) is
\begin{eqnarray}
\rho^{T_B} = \sum_{i,j,m,n}\rho_{mn}^{ij}\ket{i}\bra{j}\otimes(\ket{m}\bra{n})^T = \sum_{i,j,m,n}\rho_{mn}^{ij}\ket{i}\bra{j}\otimes\ket{n}\bra{m}
\end{eqnarray}
For 2-qubit system, $i,j,m,n \in {0,1}$.

According to PPT criterion, the partial transposition of $\rho$ has no negative eigenvalues is a necessary condition for $\rho$ to be separable. The condition is also sufficient for 2-qubit system.

Then define the entangled states we care about
\begin{eqnarray}\label{eqa:entangled_state_for_learning}
\ket{\DoubleBlochState} &=& \cos\frac{\theta}{2}\ket{00} + e^{i\phi}\sin\frac{\theta}{2}\ket{11} \\
\EnsembleForFirst &=& p\left| \DoubleBlochState  \right\rangle \left\langle \DoubleBlochState \right| + \left( {1 - p} \right)\identity/4
\end{eqnarray}
Where $0 < \theta < \pi$, $0 \le \phi \le 2\pi$ and $0 \le p \le 1$. $I$ is identity matrix.
For convenience to discuss later, we define
\begin{equation}
 \ket{0} = \begin{bmatrix}1\\0\end{bmatrix}, \quad \ket{1} = \begin{bmatrix}0\\1\end{bmatrix}
\end{equation}

And we have
\begin{equation}
\EnsembleForFirst = 
\begin{bmatrix}
p\cos^2\frac{\theta}{2} + (1 - p)\frac{\identity}{4} & 0 & 0 & pe^{-i\phi}\cos\frac{\theta}{2}\sin\frac{\theta}{2}\\
0 & 0 & 0 & 0\\
0 & 0 & 0 & 0\\
pe^{i\phi}\cos\frac{\theta}{2}\sin\frac{\theta}{2} & 0 & 0 & p\sin^2\frac{\theta}{2} + (1 - p)\frac{\identity}{4}
\end{bmatrix}
\end{equation}

\begin{equation}
\EnsembleForFirst^{T_B} = 
\begin{bmatrix}
p\cos^2\frac{\theta}{2} + (1 - p)\frac{\identity}{4} & 0 & 0 & 0\\
0 & 0 & pe^{-i\phi}\cos\frac{\theta}{2}\sin\frac{\theta}{2} & 0\\
0 & pe^{i\phi}\cos\frac{\theta}{2}\sin\frac{\theta}{2} & 0 & 0\\
0 & 0 & 0 & p\sin^2\frac{\theta}{2} + (1 - p)\frac{\identity}{4}
\end{bmatrix}
\end{equation}

The 4 eigenvalues of $\EnsembleForFirst^{T_B}$ are $\lambda_1 = p\cos^2\frac{\theta}{2} + \frac{1-p}{4}$, $\lambda_2 = p\sin^2\frac{\theta}{2} + \frac{1-p}{4}$, $\lambda_3 = \frac{1-p}{4} + p\cos\frac{\theta}{2}\sin\frac{\theta}{2}$ and
\begin{equation}\label{eqa:PPT of specific}
\lambda_{\rm min} = \frac{1-p}{4} - p\cos\frac{\theta}{2}\sin\frac{\theta}{2}
\end{equation}
$\lambda_{\rm min}$ is the minimum eigenvalue and thus $\lambda_{\rm min} < 0$ iff (if and only if) $\EnsembleForFirst$ is entangled. If $\theta = \pi/2$, the condition becomes $p > 1/3 $.

\section{\label{sec:CHSH} CHSH inequality}
CHSH (stands for John Clauser, Michael Horne, Abner Shimony, and Richard Holt) inequality, which is one of the most well-known Bell inequalities, implies the existence of local hidden variables if it is always true. In other words, the violation of CHSH inequality is a sufficient condition to ensure the state to be entangled. In this section, we apply CHSH inequality to the states of the first experiment in the main text, and show its limitation to detect entanglement.

Define CHSH operator $\Bellop$ by assembling the elements of standard CHSH inequality.
\begin{eqnarray}
\Bellop = \sigma_z\frac{\sigma_x - \sigma_z}{\sqrt{2}} - \sigma_z\frac{\sigma_x + \sigma_z}{\sqrt{2}} + \sigma_x\frac{\sigma_x - \sigma_z}{\sqrt{2}} +\sigma_x\frac{\sigma_x + \sigma_z}{\sqrt{2}}
\end{eqnarray}
Thus $-2 \le \Tr(\rho\Bellop) \le 2$, which for all separable ensemble $\rho$ predicted by CHSH inequality~\cite{Bell1989}. However, for all the entangled states, the upper bound of $\left| \Tr(\rho\Bellop) \right|$ is $2\sqrt{2}$ which was derived from Tsirelson~\cite{cirel1980quantum}.

It is easy to validate the results below.
\begin{eqnarray}
\bra{00}\Bellop\ket{00} &=& \bra{00}-\sqrt{2}\sigma_z\sigma_z\ket{00} = -\sqrt{2} \\
\bra{11}\Bellop\ket{11} &=& \bra{11}-\sqrt{2}\sigma_z\sigma_z\ket{11} = -\sqrt{2} \\
\bra{00}\Bellop\ket{11} &=& \bra{00}\sqrt{2}\sigma_x\sigma_x\ket{11} = \sqrt{2} \\
\bra{00}\Bellop\ket{11} &=& \bra{00}\sqrt{2}\sigma_x\sigma_x\ket{11} = \sqrt{2}
\end{eqnarray}

Then we obtain
\begin{eqnarray}
&&\bra{\DoubleBlochState}\Bellop\ket{\DoubleBlochState} \nonumber \\
& = & (\cos\frac{\theta}{2}\bra{00} + e^{-i\phi}\sin\frac{\theta}{2}\bra{11})\Bellop(\cos\frac{\theta}{2}\ket{00} + e^{i\phi}\sin\frac{\theta}{2}\ket{11}) \nonumber \\
& = & -\sqrt{2} + 2\sqrt{2}\cos\frac{\theta}{2}\sin\frac{\theta}{2}\cos\phi \nonumber \\
& = & \sqrt{2}(\sin\theta\cos\phi - 1) \nonumber \\
& \ge & -2\sqrt{2}
\end{eqnarray}

If $\theta = \pi/2$, $\phi = \pi$, $\ket{\DoubleBlochState}$ becomes $\frac{1}{\sqrt{2}}(\ket{00}-\ket{11})$ which maximumly violates CHSH inequality. Since $\Tr(\Bellop) = 0$, we get
\begin{equation}
\Tr(\EnsembleForFirst\Bellop) = \Tr(p\bra{\DoubleBlochState}\Bellop\ket{\DoubleBlochState}) = \sqrt{2}p(\sin\theta\cos\phi - 1) = -2\sqrt{2}p
\end{equation}
Thus for $\frac{1}{\sqrt{2}}(\ket{00}-\ket{11})$, $\EnsembleForFirst$ will not violate CHSH inequality iff $p \le 1/\sqrt{2}$. However, according to PPT we discussed in Section~\ref{sec:PPT}, all the ensembles with $p > 1/3$ are entangled. Therefore some entangled ensembles can not be detected by this CHSH inequality.

\section{\label{sec:Entanglement Witness} Entanglement Witness}
Similar to CHSH inequality, witness detection is an another method to detect entanglement. More specifically speaking, an operator $\mathcal{W}$ can be viewed as an entanglement witness iff
\begin{eqnarray}
\Tr(\rho \mathcal{W}) < 0 &&\text{   for at least 1 entangled ensemble  }\rho \\
\Tr(\rho \mathcal{W}) \ge 0 &&\text{   if  } \rho \text{ is separable}
\end{eqnarray}
We say $\mathcal{W}$ detect $\rho$ if $\Tr(\rho \mathcal{W}) < 0$. As an example, to detect an entangled state $\ket{\psi _ +} = \frac{1}{\sqrt{2}}(\ket{00} + \ket{11})$, we construct
\begin{equation}
\mathcal{W}_+ = \frac{\identity}{2} - \ket{\psi _ +}\bra{\psi _ +} = 
\frac{1}{2}\begin{bmatrix}
0 & 0 & 0 & -1\\
0 & 1 & 0 & 0\\
0 & 0 & 1 & 0\\
-1 & 0 & 0 & 0
\end{bmatrix}
\end{equation}
It is easy to validate that $\bra{\psi _ +}  \mathcal{W}_+ \ket{\psi _ +}  < 0$, thus $\ket{\psi _ +}$ is detected by $\mathcal{W}_+$. it has been shown~\cite{guhne2003experimental} that the witness must be decomposed at least in 3 measurements per party, while CHSH inequality only needs 2. 
However, for more general ensemble $\EnsembleForFirst$ with unknown $p$, $\theta$ and $\phi$, $\mathcal{W}_+$ can also detect some of entangled ensembles. Because
\begin{equation}
\Tr(\EnsembleForFirst \mathcal{W}_+)  = \frac{1-p}{4} - p\cos\frac{\theta}{2}\sin\frac{\theta}{2}\cos\phi
\end{equation}

Compared with~\ref{eqa:PPT of specific}, if $\phi = 0$ all the entangled ensembles can be detected by $\mathcal{W}_+$ perfectly whatever $p$ and $\theta$ is. However, if we do not know anything about $\phi$, this method fails.

\section{Details about our Artificial Neural Network~(ANN) model}
If not special specified, all of our ANN model consists of 3 layers - input, hidden and output layer. Every layer contains a fixed-size block of perceptrons (neurons) which connected with perceptrons in the neighbor layers.
\begin{enumerate}
	\item \textbf{input layer}. Here we denote the input layer as a vector $\vec{x}$. Our model encodes the observation results on the input layer, which includes some or all information of quantum system. Here the information - or more formally speaking - features, can be described by the expectation of observations~(every single observation or observation products) . We assume every party observes and projects the system by 2 or 3 single Pauli-matrix, which can combine to the elements of Bell inequalities or Standard State Tomography. This is why the models are called Bell-like (2 observation per party) and Tomographic (3 observation per party) predictor. TABLE ~\ref{tb} compares feature structure of our Machine learning predictors individually. 
\begin{table}\label{tb} 
	\caption{features of machine learning predictors} 
	\begin{tabular*}{15cm}{ccc}  
		\hline  
		\textbf{Name} & \textbf{Features}  & \textbf{Description} \\
		\hline 
		Tomographic Predictor  & $\bm{O^1 O^2} \cdots \bm{O^n}$, $\bm{O} \in \{\identity,\sigma_x, \sigma_y, \sigma_z \}$ & Similar to $n$-qubit state Tomography\\  
		Bell$_\text{ml}(n,3^n-1,x)$  & $\bm{O^1 O^2} \cdots \bm{O^n}$, $\bm{O} \in \{\identity,\bm{\hat{n}},\bm{\hat{n}'} \}$ & $x$ is neuron number in hidden layer\\
		&&$3^n-1$ represents feature number\\
		Bell$_\text{ml}(2, 4, x)$  & $\bm{ab}$, $\bm{ab'}$, $\bm{a'b}$, $\bm{a'b'}$ & Similar to CHSH inequality \\
		Bell$_\text{ml}(3, 4, x)$  & $\bm{abc}$, $\bm{a'b'c}$, $\bm{ab'c'}$, $\bm{a'bc'}$ & Similar to Mermin inequality \\
		Bell$_\text{ml}(3, 8, x)$  & $\bm{O^1 O^2 O^3}$, $\bm{O} \in \{\bm{\hat{n}},\bm{\hat{n}'} \}$ & Svetlichny(Double Mermin) inequality \\
		Bell$_\text{ml}(3, 12, x)$  & $\bm{ab}\identity$, $\bm{ab'}\identity$, $\bm{a'b}\identity$, $\bm{a'b'}\identity$, & Triple CHSH inequality \\
		& $\bm{a}\identity\bm{b}$, $\bm{a}\identity\bm{b'}$, $\bm{a'}\identity\bm{b}$, $\bm{a'}\identity\bm{b'}$, &\\
		& $\identity\bm{ab}$, $\identity\bm{ab'}$, $\identity\bm{a'b}$, $\identity\bm{a'b'}$ &\\
		\hline  
	\end{tabular*}
\end{table}

	\item \textbf{hidden layers}. Both hidden layers consist of perceptrons which denoted as $\vec{x}_1$.
	\begin{equation}
	\vec{x}_1 = \relu(W_1\vec{x}+\vec{w}_{01})
	\end{equation}
	where $W_1$, $\vec{w}_{01}$ are initialized uniformly and will be trained to decrease the loss function we will introduce soon. $\relu$ is a nonlinear active function for every neuron. Here ReLu~\cite{Glorot2011} are used as output function to the next layer. The definition of ReLu function $\relu$:
	\begin{equation}
	\relu ([z_1,z_2,\cdots,z_{n_e}]^T) =
    [max\{z_1,0\},max\{z_2,0\},\cdots,max\{z_{n_e},0\}]^T
	\end{equation}
    In our model, $n_e$ represents the number of neurons in the hidden layer and $\vec{x}_1 = [z_1,z_2,\cdots,z_{n_e}]^T$.
	\item \textbf{output layer}. This layer only contains 1 perceptron for 2-class identification or $m$ perceptrons for $m$-class identification, denoted as $\vec{x_2}$.
    \begin{equation}
	\vec{x}_2 = \sigma_S(W_2\vec{x_1}+\vec{w}_{02})
	\end{equation}
    Here $\sigma_S$ is sigmoid function $\sigma_S(z) = 1/(1 + e^{-z})$ for 2-class identification or softmax function
    \begin{equation}
	\sigma_S([z_1,z_2,\cdots,z_m]^T) =[\sigma_S(z_1),\sigma_S(z_2),\cdots,\sigma_S(z_m)]^T
	\end{equation}
    
    \begin{equation}
    \sigma_S(z_l) = \frac{e^{z_l}}{\sum_{j=1}^m e^{z_j}},  l=0,1,\cdots m
	\end{equation}
    for $m$-class identification.
    Every perceptron of output layer describes the probability~(defined as $p$) of every class for given quantum state predicted by model.
    
    In our work, we adopt categorical cross entropy with $m$ classes as loss function to measure the difference between model predictions and real categories~(i.e. labels). The training process of multilayer perceptron is reducing the loss function~(the sum of Eq. \ref{eq:cross_entropy} for all data, here $\hat{p}_j$ represents the probability of one state to be the $j$-th class predicted by machine, while $p_j$ is that for label. $p_j$ is either $1$ or $0$.) from the output layer by modifying the weights between neighbor layers. 
	\begin{equation}\label{eq:cross_entropy}
	H = -\sum_{j=1}^m p_j\log_2(\hat{p}_j) 
	\end{equation}
	Now we give an example (see more details in the main text) to show the calculation of cross entropy. For 4-class identification, separable states are labeled as $[1,0,0,0]^T$ while the other 3-types of entangled states are $[0,1,0,0]^T$, $[0,0,1,0]^T$ and $[0,0,0,1]^T$. if the model predicts the state as $[0.9,0.03,0.03,0.04]^T$, it actually say ''I think the state is separable with probability = 90\%." If the state is actually separable, the cross entropy is $-1 \times \log_2(0.9) - 0 \times \log_2(0.03) - 0 \times \log_2(0.03) - 0 \times \log_2(0.04)$. For 2-class identification, we only use one neuron in the output layer to encode the result.    
\end{enumerate}

Large enough data are generated to feed on our model. We divide them into 2 parts, one~($90\%-99\%$) for training the model and the other~($1\%-10\%$) for testing the accuracy. The weights of perceptrons are initially prepared on random uniform distribution. There are a variety of learning techniques for our multi-layer networks such as Stochastic gradient descent~\cite{bottou2010large}. For every iteration of training process, we shuffle training set randomly and feed the data on our model batch-by-batch. Every batch contains 32 data-label pairs. And the weights between neighbor layers are updated proportional to the gradient of loss function.
ANN architecture can be easily built from readily-available tools such as Keras~\cite{chollet2015keras}, which our model is implemented by. See Fig.~\ref{fig:Process} for more details.

As the first example in the main text, we regard CHSH inequality $\bm{a_0b_0} - \bm{a_0b_0'} + \bm{a_0'b_0} + \bm{a_0'b_0'} \ge -2$ as an entanglement classifier with $W = [1,-1,1,1]$ and $w_0 = 2$. The weights are optimized to $W = [-0.521, 0.603, -0.025, 0.016]$ and $w_0=0.373$ trained by our model (without hidden layer).

\tikzstyle{startstop} = [rectangle, rounded corners, minimum width=3cm, minimum height=1cm,text centered, draw=black, fill=red!30,text width=3cm]
\tikzstyle{io} = [trapezium, trapezium left angle=70, trapezium right angle=110, minimum width=3cm, minimum height=1cm, text centered, draw=black, fill=blue!30,text width=3cm]
\tikzstyle{process} = [rectangle, minimum width=3cm, minimum height=1cm, text centered, draw=black, fill=orange!30,text width=3cm]
\tikzstyle{decision} = [diamond, minimum width=3cm, minimum height=1cm, text centered, draw=black, fill=green!30,text width=3cm]
\tikzstyle{arrow} = [thick,->,>=stealth]

\begin{figure}
\caption{\label{fig:Process}Process of Machine Learning training}
\begin{tikzpicture}[node distance=2cm]
\node (start) [startstop] {Start};
\node (initial) [io, below of=start] {Build model and initialize weights randomly};
\node (divData) [process, below of=initial, yshift=-0.5cm] {Divide some data off for training and others for testing};
\node (testData) [process, below of=divData, yshift=-0.5cm] {Calculate accuracy of test set};
\node (dec1) [decision, right of=testData, xshift=3cm] {High enough?};
\node (pro2a) [startstop, above of=dec1, yshift=2cm] {Stop};
\node (pro2b) [process, right of=dec1, xshift=3cm] {Shuffle and divide training data into $M$ batches};
\node (calLoss) [process, below of=pro2b] {Select $i$-th batch and calculate loss function};
\node (updateW) [io, below of=calLoss, yshift=-1cm] {update weights according to gradient decent algorithm~\cite{bottou2010large}};
\node (dec2) [decision, left of=updateW, xshift=-3cm] {$i<M$?};

\draw [arrow](start) -- (initial);
\draw [arrow](initial) -- (divData);
\draw [arrow](divData) -- (testData);
\draw [arrow](testData) -- (dec1);
\draw [arrow](dec1) -- node[anchor=east] {yes} (pro2a);
\draw [arrow](dec1) -- node[anchor=south] {no} (pro2b);
\draw [arrow](pro2b) -- node[anchor=east] {$i=1$}(calLoss);
\draw [arrow](calLoss) -- (updateW);
\draw [arrow](updateW) -- (dec2);
\draw [arrow](dec2) |- node[anchor=east] {Yes, $i \coloneqq i+1$}(calLoss);
\draw [arrow](dec2) -| node[anchor=east] {No}(testData);
\end{tikzpicture}

\end{figure}
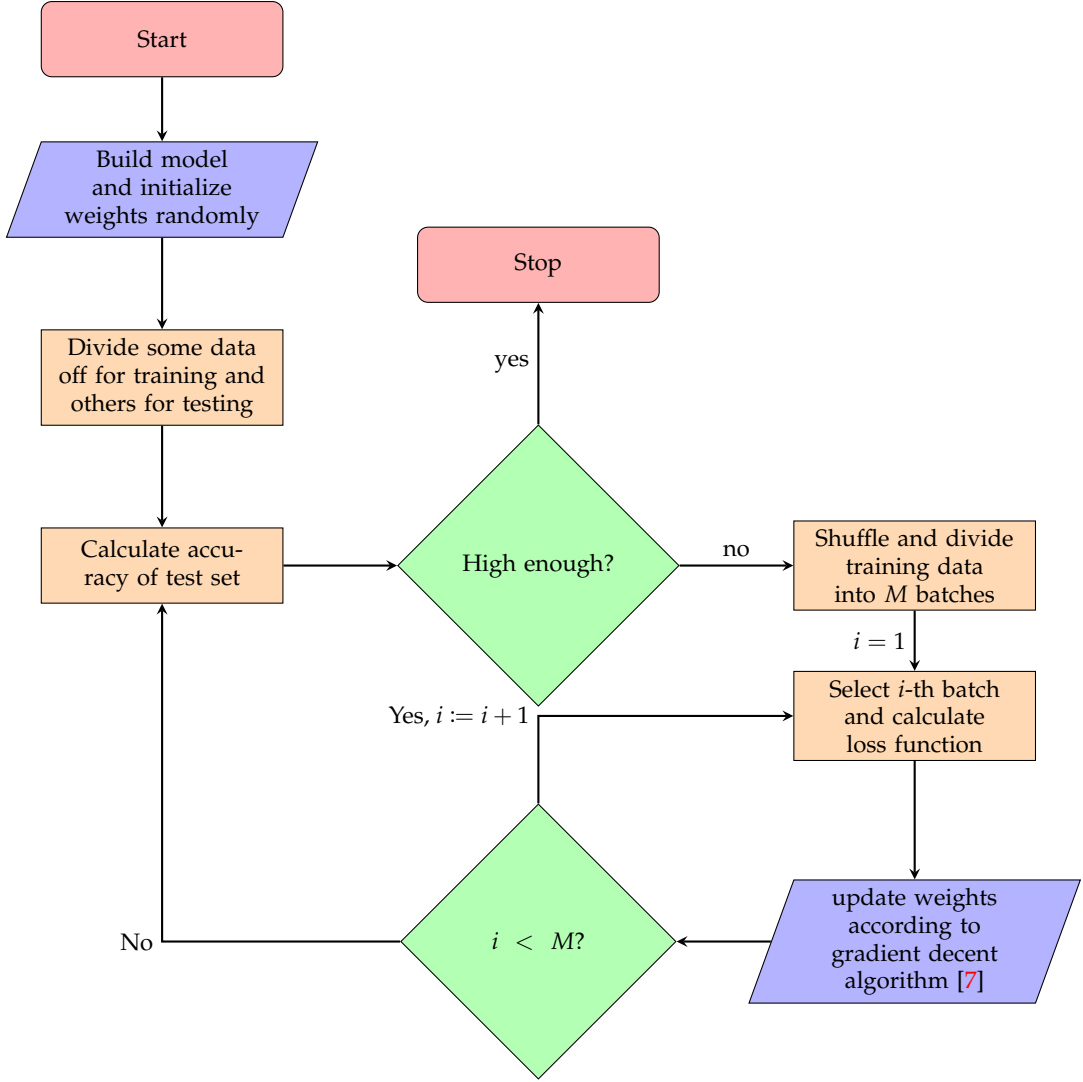

\section{methods to generate data in multi-qubit system}
In our work, All the random operators and density matrices generated accroding to Haar measure, see more details as follows or on QETLAB~\cite{qetlab}.

Our codes depend on Matlab to realize the randomness of matrices and wave functions (states). Here we just simplify and copy the kernel codes to show the method to generate these matrices and states. The size of generated matrix (state) $= \text{dim} \times \text{dim}$ ($\text{dim} \times \text{1}$). $1i = i$ is imaginary unit.




\textbf{Random Unitary Matrix(return U)}

gin = randn(dim) + 1i*randn(dim);

$\%$ QR decomposition of the Ginibre ensemble

[Q,R] = qr(gin);

$\%$ compute U from the QR decomposition

R = sign(diag(R));

R(R==0) = 1;

$\%$ protect against potentially zero diagonal entries

U = bsxfun(@times,Q,R.');

$\%$ much faster than (but equivalent to) the naive U = Q*diag(R)

\textbf{Random Density Matrix(return rho)}

gin = randn(dim) + 1i*randn(dim);

rho = gin*gin';

$\%$ Here ' is dagger.

rho = rho/trace(rho);

\textbf{Random State Vector(return v)}

v = randn(dim,1) + 1i*randn(dim,1);

v = v/norm(v); 

For example, to generate a 3-qubit pure bipartite state $\ket{\psi_{bs}}$, we can apply the code of Random State Vector with dim=2 and dim=4 and calculate their Kronecker product.